\documentstyle[bulk2]{article}

\renewcommand{\textwidth}{16.5cm}

\newcommand{\Ncollisions}{{\cal N}_{\small coll}}
\newcommand{\kt}{k_t}
\newcommand{\dc}{d_c}
\newcommand{\xt}{x_t}
\newcommand{\nt}{n_t}
\newcommand{\ntdot}{\dot{\nt}}
\newcommand{\nzero}{n_0}

\newcommand{\rhot}{\rho_t}
\newcommand{\rhotthree}{\rho_t^{(3)}}
\newcommand{\rhothree}{\rho^{(3)}}

\newcommand{\rone}{{\bf r}_1}
\newcommand{\rtwo}{{\bf r}_2}
\newcommand{\rthree}{{\bf r}_3}
\newcommand{\r}{{\bf r}}
\newcommand{\qt}{q_t}
\newcommand{\Gsep}{G^{\rm sep}}
\newcommand{\St}{S_t}
\newcommand{\mumax}{\mu^{\rm max}}

\newcommand{\Qstar}{Q^*}

\newcommand{\ta}{t_a}
\newcommand{\tstartwo}{t_{2}^{*}}
\newcommand{\tstarmany}{t_{m}^{*}}

\newcommand{\tl}{t_l}

\newcommand{\ts}{t_{\rm s}}

\newcounter{fignumber}
\begin{document}

%\bibliographystyle{europhys}

%****************************** title page *****************************************

\renewcommand{\thepage}{}

\titleben{\LARGE \bf Kinetic Regimes and Cross-Over Times in
Many-Particle Reacting Systems}

\author{\Large 
BEN O'SHAUGHNESSY\ $^{1}$ \ and \ DIMITRIOS VAVYLONIS\ $^2$ \\ 
}

\maketitle

\ \\ \newline
{\large $^1$ Department of Chemical Engineering, Columbia University,
New York, NY 10027, USA} \\
\ \\
{\large $^2$ Department of Physics, Columbia University, New York, NY
10027, USA} \\
\ \\ \ \\
\ \\ \ \\ \ \\

\pagebreak

%****************************** abstract page *****************************************

\pagenumbering{arabic}

\large

\section*{ABSTRACT}

We study kinetics of single species reactions ("$A+A \gt \emptyset$")
for general local reactivity $Q$ and dynamical exponent $z$ (rms
displacement $\xt \twid t^{1/z}$.)  For small molecules $z=2$, whilst
$z=4,8$ for certain polymer systems.  For dimensions $d$ above the
critical value $d_c=z$, kinetics are always mean field (MF).  Below
$d_c$, the density $\nt$ initially follows MF decay, $n_0 - \nt \twid
n_0^2 Q t$.  A 2-body diffusion-controlled regime follows for strongly
reactive systems ($Q>\Qstar\twid n_0^{(z-d)/d}$) with $n_0 - n_t
\approx n_0^2 \xt^d$.  For $Q<\Qstar$, MF kinetics persist, with $\nt
\twid 1/Q t$.  In all cases $\nt \approx 1/\xt^d$ at the longest
times.  Our analysis avoids decoupling approximations by instead
postulating weak physically motivated bounds on correlation functions.

\viv
\vii

PACS numbers:
  \begin{benlistdefault}
    \item [05.40.+j]
 (Fluctuation Phenomena, random processes, and Brownian
Motion)
    \item [05.70.Ln]
 (Nonequilibrium Thermodynamics, irreversible processes)   
    \item [82.35.+t]
 (Polymer reactions and Polymerization)
\end{benlistdefault}

\pagebreak

%****************************** PAPER  *****************************************

The kinetics of reactions between diffusing particles are anomalous in
low spatial dimensions
\citeben{kotominkuzovkov:book_short,kangredner:segregation}.  Mean field
(MF) theory, according to which the reaction rate is proportional to
the product of reactant densities, is only valid for dimensions $d$
above a critical value $\dc$.  In the most fundamental problem of
single-species reactions into inert products, $A+A\gt \emptyset$, for
$d<\dc=2$ the density decays asymptotically as $1/t^{d/2}$,
independently of the magnitude of the local chemical reactivity $Q$
and initial reactant density.  That is, in low dimensions the $1/Qt$
decay predicted by MF theory is invalid.
This is supported by numerical simulations
\citeben{kangredner:segregation,toussaintwilczek:segregation,meakinstanley:aplusb_fractal,%
argyrakis:aplusb_crossovers} and renormalization group studies
\citeben{lee:aa_rg}.  For the case of ``infinitely'' reactive
particles for which $Q = \infty$ (probability of reaction per
collision unity), rigorous bounds
\citeben{bramsonlebowitz:aplusb_europhys}
on the asymptotic decay of density have been derived and in one
dimension exact solutions exist
\citeben{torneymcconnel:aplusa_exact_1d,lushnikov:aplusa_exact_1d,%
spouge:aplusa_exact_1d,doeringbenavraham:aplusa_exact_1d}.  Many other
workers have employed approximate methods starting from the hierarchy
of coupled dynamical equations for correlation functions of all orders
which is then truncated by expressing higher order correlation
functions in terms of lower order correlations
\citeben{kotominkuzovkov:book_short,lindenberg:aplusa_jphyschem}.

In this work we address the two major aspects of single species
reaction kinetics which remain poorly understood.  (1) Although it is
known that at the shortest timescales MF kinetics apply
\citeben{privman:aplusa_finite_Q,zhongbenavraham:aplusa_finite_Q}, the crossover
from the short time MF kinetics to the asymptotic $1/t^{d/2}$ behavior
has not been established.  Some groups on the basis of numerical
simulations
\citeben{braunstein:aplus_finite_Q,shikopelman:reactions_finite_Q,%
martinbraunstein:aplusa_finite_Q_zphys} have claimed an
``intermediate'' time regime during which density decays according to
a non-universal $Q$-dependent power law.  (2) No systematic theory has
been able to predict reaction kinetics for general $Q$ 
\citeben{lee:aa_rg,privman:aplusa_finite_Q,zhongbenavraham:aplusa_finite_Q}
across all time regimes.  As a result numerical simulations have been
fitted to empirical rate laws
\citeben{hoyuelosmartin:aplusa_finite_Q_langmuir,hoyuelosmartin:aplusa_finite_Q_pre}.

The present work will also consider general dynamics.  The above
results were for reactions between particles obeying simple Fickian
diffusion for which the dynamical exponent $z$ describing rms
displacement $\xt$ as a function of time, $\xt \approx a
(t/\ta)^{1/z}$, is equal to 2.  Here $\ta$ is the time corresponding
to diffusion distance of order the particle size $a$.  For reactive
groups attached to long polymer chains $z$ can also be 4 or 8,
depending on time and degree of entanglement
\citeben{doiedwards:book}.  In these cases the critical dimension is
$\dc=z$ \citeben{gennes:polreactionsiandii} below which the asymptotic
density decay is $1/\xt^d$ \citeben{oshanin:review}.

In this letter we give a complete description of the sequence of
kinetic regimes for the single species reaction kinetics for arbitrary
values of dynamical exponent $z$ (where $z$ is assumed a fixed number
independent of dimension).  Results are derived after postulating
simple physical bounds on correlation functions without the need to
resort to ad-hoc decoupling approximations.  In agreement with
previous studies, we find that below the critical dimension $\dc=z$
the density $\nt$ decays as $1/\xt^d$ at long times.  We find two
possible kinetic sequences for $d<\dc$, depending on the magnitude of
$Q$ with respect to a marginal value $\Qstar$.  For $Q<\Qstar$
(``Weak'' systems) the crossover to $\nt \approx 1/\xt^d$ occurs after
a $1/t$ MF regime.  For $Q>\Qstar$ (``Strong'' systems) the crossover
to $1/\xt^d$ occurs before the $1/t$ regime has developed.  In this
case a short time regime exists during which the number of reactions
is proportional to the reactant exploration volume, $\nzero - \nt
\approx \nzero^2 \xt^d$.

We begin by noting that the average reaction rate $\ntdot \equiv
{d \over dt} \nt$ at point $\rone$ is proportional
\citeben{doi:reaction_secondquant1and2} to to the 2-body correlation
function $\rhot(\rone,\rtwo)$ evaluated at $\rone=\rtwo$.  Due to
translational invariance $\ntdot = -\lambda \rhot(0,0)$ where $\lambda
\equiv Q a^d$.  Now the dynamical equation for $\rho$ involves the
3-body correlation function $\rhothree$.  Following Doi
\citeben{doi:reaction_secondquant1and2} the exact equation is 
\pagebreak
%_______________________________________________________________________
                                                \begin{eqarray}{pepper}
\rhot(\rone,\rtwo) &=& \nzero^2 
- \lambda \int d\rone' \, d\rtwo' \int_0^t dt' \,
        G_{t-t'}(\rone,\rtwo,\rone',\rtwo')\, 
         \rho_{t'}(\rone',\rtwo') \delta(\rone'-\rtwo')
                                                        \drop
&-& \lambda \int d\rone' \, d\rtwo' \, d\rthree' \int_0^t dt' \,
        G_{t-t'}(\rone,\rtwo,\rone',\rtwo') \, 
        \rho_{t'}^{(3)}(\rone',\rtwo',\rthree')
        \curly{\delta(\rone'-\rthree') + \delta(\rtwo' - \rthree')}
\comma 
\drop
                                                          \end{eqarray}
%-----------------------------------------------------------------------
where $G_t(\rone,\rtwo,\rone',\rtwo')$ is the net weighting for two
particles to arrive at $\rone,\rtwo$ given starting points
$\rone',\rtwo'$, in the absence of reactions.  The sink terms on the
right hand side of eq. \eqref{pepper} describe the three ways in which
reactions diminish $\rhot$ from its initial value $\nzero^2$.  The
first {\em two-body} sink term subtracts off pairs which failed to
reach $\rone,\rtwo$ because their members reacted with one another at
$\rone'$ at time $t'$.  The remaining two sink terms subtract off
pairs which would be at $\rone,\rtwo$ but only one member of which
reacted at time $t'$ at location $\rthree'$.  Such a reaction involves
a third particle, weighted by the appropriate 3-body correlation
function.  In eq. \eqref{pepper} we used a $\delta$ function as a
reactive sink which is a coarse-grained description of the reaction
process over a scale of order the particle size $a$.

Notice that eq. \eqref{pepper} is not closed in terms of $\rhot$ since
it involves the unknown $\rhotthree$.  It is in fact impossible to
write a closed exact equation for $\rhot$ since correlation functions
of all orders are coupled in an infinite hierarchy of dynamical
equations \citeben{doi:reaction_secondquant1and2}.  This complication
may be resolved by approximating 3-body correlations as products of
lower order correlation functions
\citeben{kotominkuzovkov:book_short}.  However we are able to write a
closed equation for $\rhot$ after assuming much less restrictive
bounds on the magnitude of correlation functions.  First we transform
eq. \eqref{pepper} to an expression for the function $\qt(\r,0) \equiv
\rhot(\r,0)/\nt^2$.  Notice that $\qt(0,0) = \kt/\lambda$ where the
rate constant $\kt$ is defined by $\ntdot=- \kt
\nt^2$.  We find that eq. \eqref{pepper} is transformed into
%_______________________________________________________________________
                                                \begin{eq}{alien}
\qt(\r,0) = 
 1 -  \int_0^t dt'\, \Gsep_{t-t'}(\r,0) \, k_{t'} \ + \ \varphi_t(\r) 
\comma
                                                          \end{eq}
%-----------------------------------------------------------------------
where $\Gsep_t(\rone,\rtwo) \equiv \int d\rtwo'\,
G_t(\rone',\rone'+\rone,\rtwo',\rtwo'+\rtwo)$ is the probability
density two particles are separated by $\rone$ at time $t$ given
initial separation $\rtwo$.  The function $\varphi_t$ is defined by
%_______________________________________________________________________
                                                \begin{eq}{camel}
\varphi_t (\r) \equiv
\int_0^t dt' \int d\r' \, \Gsep_{t-t'}(\r,\r')\, k_{t'}\, \mu_{t'}(\r')
\comma \gap
\mu_{t}(\r) \, \equiv \, 2 \curly{
\rhot(\r|0) - \rhotthree(\r|0,0) 
} \period \drop
                                                                \end{eq}
%-----------------------------------------------------------------------
Here $\rhot(\r|0) \equiv {\rhot(\r,0)/ \nt}$ and $\rhotthree(\r|0,0) \equiv
\rhotthree(\r,0,0)/ \rhot(0,0)$ are the conditional 
densities at $\r$, given one and two particles, respectively, at the
origin.

Now we would like to solve eq. \eqref{alien} for $\qt(0,0)$. This
requires information on the properties of $\varphi_t(0)$ which
involves unknown 2-body and 3-body conditional densities.  To proceed,
let us now make the following assumption: We assume that the more
particles placed at the origin, the lower the conditional density.
Chemical reactivity can only induce anticorrelations.  Thus:
%_______________________________________________________________________
                                                \begin{eq}{assumption}
\rhotthree(\r|0,0)\, \leq\, \rhot(\r|0)\, \leq \nt
 \gap \mbox{(assumption)}
\period
                                                                \end{eq}
%-----------------------------------------------------------------------
Eq. \eqref{assumption} implies the following constraints on $\mu$:
%_______________________________________________________________________
                                                \begin{eq}{constraints}
0 \, \leq \,\mu_t (\r)\, \leq \, 2 \nt 
\comma \gap
\int_0^t dt' k_{t'} \, \ge \, \int_0^t dt' \int d\r' \, k_{t'} \,
\mu_{t'} (\r') \period
                                                                \end{eq}
%-----------------------------------------------------------------------
The first constraint immediately follows from eq. \eqref{camel} while
the second is obtained by requiring that the magnitude of the second term
on the right hand side of eq. \eqref{alien} exceeds that of the 3rd term
(since $\qt\le 1$) and then integrating over $\r$.

On the strength of the above constraints, eq. \eqref{constraints}, we
will argue below that the solution of eq. \eqref{alien} for
$\qt(0,0)$, but with the term $\varphi_t(0)$ deleted, gives the
correct power law solution for $\kt$ to within a constant prefactor.
Expressing $\qt(0,0)$ in terms of $\kt$, setting $\r=0$ and deleting
$\varphi_t(0)$, eq.
\eqref{alien} becomes
%_______________________________________________________________________
                                                \begin{eq}{clock}
\kt = \lambda - \lambda \int_0^t dt' S_{t-t'} k_{t'} \comma
                                                                \end{eq}
%-----------------------------------------------------------------------
where we have introduced the return probability $\St \equiv
\Gsep_t(0,0) \approx 1/\xt^d$.  The validity of this approximation is
justified by self-consistent arguments whose outline is as follows
(full details will be published elsewhere
\citeben{ben:bendimitris:tobepublished}).  If one accepts eq.
\eqref{clock} then one obtains a sequence of power law regimes in time
for $\kt$ and $\nt$ (see below).  Using these solutions in eq.
\eqref{constraints} and in the expression for $\varphi_t(0)$ (eq.
\eqref{camel}), we have determined the function $\mumax_{t'}(\r')$
which maximizes $\varphi_t(0)$ (for a given time $t$) subject to the
constraints of eq. \eqref{constraints}.  This in turn implies an upper
bound $\varphi_t^{\rm max}(0)$ on $\varphi_t(0)$.  Now for short
times, we find that this bound is much less than unity: hence
$\varphi_t(0)$ may be deleted in eq. \eqref{alien} without error and
even prefactors are expected to be correct.  Meanwhile, for long times
($n_t\approx 1/\xt^d$) we find $\varphi_t^{\rm max}(0) = A$, where $A$
is a constant of order unity.  We then make a 2nd simple assumption:
we assume $\kt$ remains a power law at long times.  We have shown that
the deletion of $\varphi_t(0)$ from eq. \eqref{alien} is then valid
insofar as it will generate the correct power laws for $\kt$ (albeit
with possibly incorrect prefactors).

It is now straightforward to solve eq. \eqref{clock} for $\kt$.  For
$d>z$ the integral term in eq. \eqref{clock} is negligible and one
recovers the expected MF rate constant $k \approx \lambda$, true for
all times.  For $d \le z$ one finds two regimes: for $t<\tstartwo$
(see definition eq. \eqref{timescales} below) the integral term in eq.
\eqref{clock} is small compared to $\lambda$ and can be neglected.  For
$t>\tstartwo$, the solution is obtained by balancing the integral term
with $\lambda$.  One finds
%_______________________________________________________________________
                                                \begin{eq}{monkey}
\kt \approx \casesbracketsshortii{\lambda}{t \ll \tstartwo}
                             {\xt^d/t}{\tstartwo \ll t}
\ \ \  (d<z)
\comma \gap
\kt \approx \casesbracketsshortii{\lambda}{t \ll \tstartwo}
                             {a^d/ [\ta \ln(t/\ta)]}{\tstartwo \ll t}
\ \ \  (d=z)
\comma
                                                                \end{eq}
%-----------------------------------------------------------------------
as can be verified by direct substitution in eq. \eqref{clock}.
Substituting $\kt$ in $\ntdot = -\kt \nt^2$, it is easy to show that
for low dimensions two possible sequences of reaction kinetics may
occur, depending on the magnitude of $Q$:
%_______________________________________________________________________
                                                \begin{eq}{decay}
\nt \, \approx\, \casesbracketsii
{\nzero - \lambda \nzero^2 t\
\stackrel{\tstartwo}{\ggt} \
\nzero - \nzero^2 \xt^d\
\stackrel{\tl}{\ggt} \
1/\xt^d\ }
{Q>\Qstar,\ \mbox{Strong}}
{\nzero - \lambda \nzero^2 t\
\stackrel{\tstarmany}{\ggt} \
\ \ \ \ 1/\lambda t \ \ \ \ 
\stackrel{\tstartwo}{\ggt} \
1/\xt^d \ }
{Q<\Qstar,\ \mbox{Weak}}
\gap (d<z)
                                                                \end{eq}
%-----------------------------------------------------------------------
where 
%_______________________________________________________________________
                                                \begin{eq}{timescales}
{\tstartwo \over \ta} \equiv (Q \ta)^{z/(d-z)} 
\comma \gap
{\tl \over \ta} \equiv (\nzero a^d)^{-z/d}
\comma \gap 
\tstarmany \equiv {1 \over \lambda \nzero}
\comma \gap
\Qstar \ta \equiv (\nzero a^d)^{(z-d)/d} 
%\gap (d<z)
\period
                                                                \end{eq}
%-----------------------------------------------------------------------
For the marginal case ($d=z$) one simply replaces $\xt^d$ in eq.
\eqref{decay} by $a^d (t/\ta)/\ln(t/\ta)$.  The timescales $\tstartwo$
and $\tl$ are then modified to $\tstartwo \equiv \ta e^{1/(Q \ta)}$,
$\tl
\equiv \ta (\nzero a^d)^{-1} \ln[1/(\nzero a^d)]$, while $\Qstar
\ta \equiv 1/\ln(\tl/\ta)$.

These results for $d<z$ have a simple physical interpretation (see fig.
\ref{kinetic_regimes}).  Now if one were to assume MF theory were
valid, \ie that particles are distributed at all times as in equilibrium
and hence $\ntdot = - \lambda \nt^2$, the density would decay as $\nzero -
\lambda \nzero^2 t$ initially, crossing over to $1/\lambda t$ for $t
\gg \tstarmany$.  Thus $\tstarmany$ would be the timescale for a particle to
react with the mean reaction field supplied by the other reactants.
For dimensions smaller than the critical value $\dc=z$, however,
reactions induce nonequilibrium correlations and MF kinetics break
down beyond a certain timescale. Consider a
pair which happened to be initially within diffusive range,
\ie within $\xt$, as in fig. \ref{kinetic_regimes}(a).  
Since each reactant explores a volume of order $\xt^d$ approximately
uniformly, the number of collisions
$\Ncollisions$ after $t/\ta$ ``steps'' increases roughly as
$(t/\ta)(a^d/\xt^d) \twid t^{1-d/z}$.  Thus for $d<\dc$ the net
reaction probability $Q \ta \Ncollisions$ is an increasing function of
time which becomes of order unity at the timescale $\tstartwo$ of eq.
\eqref{timescales}.  For $t>\tstartwo$ a depletion hole of size $\xt$
thus develops in the 2-body correlation function, invalidating MF
theory.

For $d<\dc$, the sequence of kinetic regimes after $\tstartwo$ depends
on the relative magnitudes of $\tstartwo$ and $\tstarmany$.  Another
relevant timescale is $\tl$, namely the time to diffuse a distance of
order the typical initial particle separation.  One can easily show
that the magnitude of $\tstarmany$ always lies in between those of
$\tstartwo$ and $\tl$.  The condition $\tstartwo=\tstarmany=\tl$
defines the critical reactivity $\Qstar$ in eq. \eqref{timescales}.
\underline{\bf Case 1:}  Strongly reactive species, $Q>\Qstar$ (or
equivalently $\tstartwo<\tstarmany<\tl$).  For $\tstartwo<t<\tl$
reactions are due to the few isolated pairs which happened to be
initially within diffusive range (fig. \ref{kinetic_regimes}(a)).
Since most of these pairs will have reacted by $t$, the number of
reactions per unit volume during this time regime is proportional to the
number of pairs per unit volume initially within $\xt$, $\nzero^2
\xt^d$.  Now for times $t>\tl$, a region of volume $\xt^d$ contained
initially several particles.  Since only of order one of these
particles could have survived by $t$, this implies a $1/\xt^d$ density
decay.  Notice that $\tstarmany$ is irrelevant in strong systems.
\underline{\bf Case 2:} Weakly reactive species, $Q<\Qstar$ (or
$\tl<\tstarmany<\tstartwo$). Now when nonequilibrium correlations develop
after $\tstartwo$ the $1/\lambda t$ MF kinetics have already
developed.  Since $\tstartwo>\tl$, the same reasoning as in the
strong case for $t>\tl$ implies that a direct crossover occurs
after $\tstartwo$ to $1/\xt^d$ decay.

To summarize, we argued that the time dependent rate constant $\kt$ in
single-species reaction kinetics is given by a Smoluchowski type
expression (eq. \eqref{clock}).  Equivalent expressions have been the
starting point of previous works
\citeben{torneymcconnel:aplusa_2d_smoluchowski,zumofen:aplusa_smoluchowski,oshanin:review}.
Here we have justified eq. \eqref{clock} based on simple bounds on
correlation functions.  We argued that the solution for the rate
constant below the critical dimension will then be correct at short
times, but there is evidence that at long times (during the $1/\xt^d$
density regime) the actual numerical prefactor will be larger than
that predicted by eq. \eqref{clock}.  Indeed, for infinitely reactive
particles in one dimension for $z=2$, Torney and McConnell
\citeben{torneymcconnel:aplusa_exact_1d} have proved that $\kt$ as
determined by eq. \eqref{clock} is correct at short times, but is
smaller by a factor of $\pi/2$ from its actual value at long times.

Our result for the density decay (eq. \eqref{decay}) does not give
evidence for a $Q$-dependent power law decay at intermediate times as
suggested in refs.
\citenum{braunstein:aplus_finite_Q,shikopelman:reactions_finite_Q,%
martinbraunstein:aplusa_finite_Q_zphys} for $z=2$.  We suggest these
workers may have been observing intermediate kinetics between the
$1/t$ and $1/t^{d/2}$ regimes of eq. \eqref{decay}.

We remark that the results here have been for general diffusion
dynamics described by the dynamical exponent $z$.  They apply for both
subdiffusion ($z>2$) and superdiffusion ($z<2$) provided that
reactants diffuse independently of one another, and that $\xt$ and
mean collision time $\ta$ are well-defined.  Implicitly, we have
assumed that all moments of displacement are characterized by the same
scale, $\xt$.  Systems in
which reactants perform L\'{e}vy walks, for which the rms
displacement does not exist
\citeben{zumofenklafter:react_levy_pre,oliva:aplusa_levy_analytical},
are not covered.  The results do not apply also to systems in which
reactants perform random walks for which the probability distribution
for the time between successive steps has a power law long-time tail
and hence $\ta$ does not exist
\citeben{blumen:react_anomalous_fract,alemany:aplusa_ctrw}.  In order
to describe systems of the latter type, eq. \eqref{pepper} should be
modified to include an extra integration over all possible
collision times.

To specialize to the most widely studied case of small Fickian
molecules one sets $z=2$.  For example, at the critical dimension
$d=2$ we predict a short time diffusion-controlled regime for high
reactivity $(Q>Q^*)$ in which $\nzero-\nt\twid \nzero^2 t/\ln t$, and
we predict that in all cases the long time decay is $\nt\twid \ln
t/t$.

Finally, it is worth discussing how the above results may generalize
to the two-species reaction problem, $A+B \gt \emptyset$, for which
segregation of reactants into A-rich and B-rich regions occurs at long
times in dimensions $d<2z$.  This leads to an asymptotic decay $\nt
\approx (\nzero/\xt^d)^{1/2}$
\citeben{kangredner:segregation,toussaintwilczek:segregation}.
For $d<z$, assuming that these kinetics apply for times longer than a
timescale $\ts$, we may estimate the magnitude of $\ts$ by assuming
that eq. \eqref{decay} applies for $t<\ts$ and demanding continuity in
$\nt$.  Thus one finds $\ts \approx \tl$ for strong systems, while for
weak systems $\ts \approx \ta [\nzero a^d (Q \ta)^2]^{z/(d-2z)}$ which
satisfies $\tstarmany<\ts<\tstartwo$.  This would suggest that at
least for weak systems the $1/\xt^d$ regime does not arise in $A+B \gt
\emptyset$.  Indeed, this is consistent with
the numerical simulations of ref.
\citenum{kangredner:segregation}, where no $1/\xt^d$
regime was found for the case $z=2$, $d=1$.  These workers observed
that for sufficiently short times the density remained close to
$\nzero$.  For high $Q$ values a crossover to $t^{-1/4}$ occurred at
$\tl$.  For small $Q$, they found that the $t^{-1/4}$ regime was
preceded by a $1/t$ regime during $\tstarmany < t <\ts$.  (We
interpret the onset time here to be $\tstarmany$, not $\tl/(Q\ta)$ as
interpreted by the authors of ref.
\citenum{kangredner:segregation}.  The numerical values of these two
timescales happen to be very close to one another in this simulation.)
We hope future numerical simulations as well as experiments on
reacting polymers where novel values of $z$ are realized will test the
validity of the above theoretical predictions.

\centerline{***}

This work was supported by the National Science Foundation, grant no.
DMR-9403566.

%****************************** BIBLIOGRAPHY *****************************************

\pagebreak

%\bibliography{general,polreaction,allreaction,polymerization,polgeneral,ben,rgcritical,radical,polinterface,allreaction_dimitris}

\pagebreak

%*************************************************************************************
%*************************************************************************************
%*************************** BEGIN FIGURE CAPTIONS ***********************************
%*************************************************************************************

                     \begin{thefigures}{99}

%*************************************************************************************

\figitem{kinetic_regimes}

Snapshots of particle distributions during various kinetic regimes for
$d<z$.  Rms displacement is $\xt$ and $l$ is typical initial
separation among reactants.  (a) Times $t\ll\tl$.  Reactions are due
to the few isolated pairs which happened to be initially close enough
such that their exploration volumes (indicated by dashed lines)
overlap by $t$.  For strong systems most of these pairs will have
reacted by $\tstartwo$, leading to $\nzero-\nt \approx \nzero^2
\xt^d$.  (b) The situation at time $t=\tl$.  A region of size $\xt^d$ contained of
order one particle initially.  For strong systems reaction kinetics
are already diffusion-controlled and a crossover occurs to $\nt
\approx 1/\xt^d$.  (c) Weak systems, $\tl \ll t\ll \tstartwo$.  For weak
systems MF kinetics persist beyond $\tl$.  Many reactants now exist
within the exploration volume of a given particle.  Reaction with the
mean density field occurs at a timescale $\tstarmany$ leading to $\nt
\twid 1/Q t$ for $\tstarmany
\ll t \ll \tstartwo$.  (d) For both weak and strong systems, at 
sufficiently long times only of
order one particle survives within a volume of size $\xt^d$, implying
$\nt \approx 1/\xt^d$.

%*************************************************************************************

                     \end{thefigures}

%*************************************************************************************
%*************************************************************************************
%***************************** END FIGURE CAPTIONS ***********************************
%*************************************************************************************

\pagebreak

%
%
%********************************************************************************
%********************************************************************************
%
%                              FIGURES FOR LETTER VERSION
%
%********************************************************************************
%********************************************************************************
% NOTE FIGURES ARE EXPECTED IN PORTRAIT MODE IN THE FOLLOWING
%

\begin{figure}[t]

\epsfxsize=\textwidth \epsffile{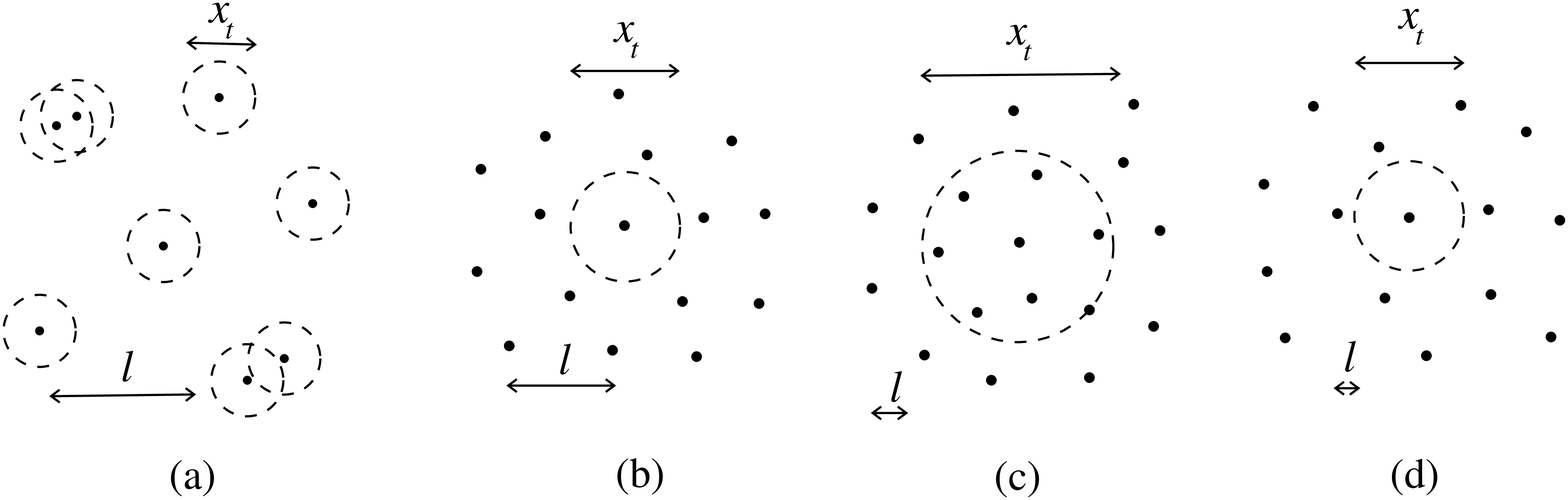}

\end{figure}

\mbox{\ }

\vfill

\addtocounter{fignumber}{1}
\mbox{\ } \hfill {\huge Fig.\@ \thefignumber} 

\pagebreak
%*******************************************************************************

\end{document}